\documentclass[a4paper]{jpconf}
\usepackage{graphicx,amssymb}

\begin{document}

\title{Sources of UHECRs in view of the TUS and JEM-EUSO experiments}

\author{N N Kalmykov, B A Khrenov, G V Kulikov and M Yu Zotov}

\address{Lomonosov Moscow State University, Skobeltsyn Institute of
Nuclear Physics,\\ Moscow, Russia, 119234}

\ead{zotov@eas.sinp.msu.ru}

\begin{abstract}
The origin of ultra-high-energy cosmic rays (UHECRs) is one of the most
intriguing problems of modern cosmic ray physics. We briefly review the
main astrophysical models of their origin and the forthcoming orbital
experiments TUS and JEM-EUSO, and discuss how the new data can help one
solve the long-standing puzzle.
\end{abstract}

In 1984, Hillas suggested a simple necessary condition for an
astrophysical object to be able to accelerate charged particles to ultra-high
energies~\cite{hillas-1984}. The criterion is based on the demand that
the Larmor radius of a particle of charge~$Z$ in a magnetic field~$B$
should not exceed the size~$R$ of the acceleration region. In its
simplest form, this boils down to the following estimate of the maximum
energy that can be obtained by a particle: $E_{\max}\lesssim\beta Z B R$,
where~$B$ is expressed in units of $\mu$G, $R$ is in kpc, energy
is in the units of EeV, and~$\beta$ represents the velocity of the
accelerating shock wave or the efficiency of the accelerator. 
The criterion can be illustrated by a now famous Hillas diagram.
Different estimates of parameters of cosmic accelerators result in
slightly different Hillas diagrams, compare, for example, those given
in~\cite{pt-troi-2010,kotera-olinto-2011,selvon-stanev-2011,sigl-2012}.
Still, almost all authors select the following four major classes
of possible sources of UHECRs:
large-scale shocks due to merging galaxies or clusters of galaxies,
the core and jets of active galactic nuclei (AGN),
newly-born neutron stars and other similar objects,
and gamma-ray bursts.


\textbf{Cluster accretion shocks} are the largest systems in the known
universe that satisfy the Hillas criterion.
They have been considered as possible sources of UHECRs by various
authors, see, e.g.,~\cite{norman_etal-1995,kang_etal-1996} for the
early discussions.
In particular, Kang et~al.~\cite{kang_etal-1996} suggested that protons
can be accelerated by these accretion shocks up to energies
$\sim6\times10^{19}$~eV assuming acceleration is performed via the
first-order Fermi process and provided the mean magnetic field strength
in the region around the shocks is $\gtrsim1\mu$G.
Later calculations that were based on the framework of diffusive shock
acceleration (DSA) and included energy losses due to interactions of protons
with photons of the CMBR lead to somehow more modest estimates of the
maximum achievable energy but the conclusion remained intact: under
certain conditions, galaxy clusters are able to accelerate protons and
heavier nuclei to ultra-high energies~\cite{vannoni_etal-2009}.
It is worth mentioning that particles accelerated via DSA in shocks
that occur during cluster mergers, together with magnetic fields,
emit synchrotron radiation and may form so-called radio relics.
One such relic with the size of 2~Mpc and highly aligned magnetic
fields has been detected recently in the northern outskirts of the galaxy
cluster CIZA J2242.8+5301 ($z=0.1921$) providing evidence that DSA can operate
on scales much larger than in supernova remnants and that shocks in
galaxy clusters are capable of producing UHECRs~\cite{van_weeren_etal-2010}.


\textbf{Active galactic nuclei} are probably the most popular
astrophysical objects potentially capable to produce and accelerate
particles up to the highest energies.
They are composed of an accretion disk around a central super-massive
black hole and are sometimes associated with jets terminating in lobes
(or hot spots) that can be detected in radio, see,
e.g.,~\cite{kotera-olinto-2011}.
AGN can be divided into two groups: radio loud objects with jets
and radio quiet ones with no prominent radio emission or jets.
AGN jets have dimensions of the order of a parsec with magnetic
fields of the order of a few Gauss and can in principle accelerate
protons up to tens of EeV.
In their turn, AGN cores with a magnetic field $\sim10^3$~G and the size
of a few $10^{-5}$~pc can reach approximately the same
energy~\cite{selvon-stanev-2011}.
However, these maxima are unlikely to be reached because of
adiabatic losses, energy losses due to synchrotron radiation
and inverse Compton processes, as well as photodisintegration of
heavy nuclei thus reducing the maximum achieved energy to
only a fraction of EeV~\cite{bhat-sigl-2000}.
To get around the problem, the acceleration site must be away
from the AGN center and in a region with a lower radiation density.
Among radio loud galaxies, the Fanaroff--Riley galaxies of class~II
are of special interest because they combine a powerful engine
and relativistic blast wave together with a relatively scarce
environment.

AGN as possible sources of UHECRs attract special attention after
the Pierre Auger Observatory (PAO) reported correlation of their
higher energy events with the nearby AGN (distances within 75~Mpc)
\cite{PAO-agn-2007,PAO-agn-2008,PAO-agn-2010}.
The largest excess of arrival directions relative to isotropic
expectations was found around the position of Cen~A, the nearest
radio loud AGN located at a distance of 3.8~Mpc from the Solar system.
Still, this is not a proof that AGN are sources of UHECRs.
A similar analysis performed by the HiRes experiment did not
reveal any deviation from the isotropic distribution~\cite{hires-agn-2008}.
One should note though that an analysis of anisotropy of arrival
directions of UHECRs is not a trivial task and different conclusions
can be obtained with the same data, see, e.g.,~\cite{rubtsov_etal-2012}.


\textbf{Pulsars} attracted attention as possible accelerators of UHECRs
soon after they were identified with rotating neutron stars possessing
huge magnetic fields.
It was suggested that protons can obtain energies
up to $10^{21}$~eV by riding in the outgoing strong wave field
beyond the light cylinder of a pulsar~\cite{gunn-ost-1969}.
It was soon demonstrated though that maximum energies to which
particles are accelerated may be considerably lower because
even a small contamination of plasma in the waves would destroy
the necessary phase locking~\cite{kegel-1971}.

An interest to pulsars as possible accelerators received another impetus
when it was demonstrated that young strongly magnetized neutron stars
with short enough initial spin periods are able to
accelerate iron nuclei to energies $\gtrsim10^{20}$~eV through
relativistic MHD winds~\cite{blasi_etal-2000}.
Recent calculations~\cite{newPSRs-olinto-2012} show
that, at early times, when protons can be accelerated to energies
$E>10^{20}$~eV, the young supernova shell tends to prevent their escape.
In contrast, because of their higher charge, iron-peaked nuclei are still
accelerated to the highest observed energies at later times, when the
envelope has become thin enough to allow their escape. Ultra-high energy
iron nuclei escape newly-born pulsars with millisecond periods and dipole
magnetic fields of $\sim10^{12}$--$10^{13}$~G, embedded in core-collapse
supernovae.
It is remarkable that due to the production of secondary nucleons, the
envelope crossing leads to a transition from light to heavy composition
at a few EeV, as was recently found by the
PAO~\cite{PAO-Xmax-2010,PAO-mass-2012} (provided that the models used
give a fairly proper description of the physical processes at these
energies).
According to ~\cite{newPSRs-olinto-2012}, the escape also results in a
softer spectral slope than that initially injected via unipolar
induction, which allows for a good fit to the observed UHECR spectrum.


\textbf{Gamma-ray bursts} were independently suggested as possible
candidate sources of UHE particles
in~\cite{waxman-1995,vietri-1995,milgrom-usov-1995}.
Waxman and Vietri demonstrated basing on the fireball model by
M\'esz\'aros and Rees~\cite{mesz-rees-1994} that protons can be
accelerated to energies $\sim10^{20}$~eV by the conventional Fermi
mechanism at highly relativistic shocks. The model predicted a mostly
isotropic, time-independent distribution of arrival directions of UHECRs
because so are GRBs, and correctly predicted the total flux of UHECRs at
Earth by an order of magnitude.

In 1997, Waxman and Bahcall found that a large fraction ($\ge10$\%) of
the energy of a fireball producing a gamma-ray burst is expected to be
converted by photomeson production to a burst of $\sim10^{14}$~eV
neutrinos~\cite{waxman-bahcall-1997}.
More than this, it was demonstrated that a kilometer-scale neutrino detector
would observe at least several tens of events per year correlated
with GRBs in case the GRB model of UHECR origin is correct.
Since then, the model received much attention even though some authors
remained sceptical about it.
One of the challenges for the GRB model relates to the trend toward
a heavy composition at the highest energies found by the
PAO~\cite{PAO-Xmax-2010,PAO-mass-2012}.

Doubts in the validity of the model became stronger after
the IceCube collaboration published their latest results on the search
for neutrinos associated with GRBs~\cite{icecube-2012-nu},
confirming the earlier conclusions~\cite{icecube-nus-2011a,icecube-nus-2011b}.
Namely, the IceCube collaboration studied the data obtained while the
telescope was under construction using the 40- and 59-string configurations
of the detector, which took data from April 2008 to May 2009 and from May 
2009 until May 2010, respectively.
Totally, around 300 GRBs were observed and reported
via the GRB Coordinates Network during the two data taking periods
and included in the study.
Two analyses of the IceCube data were performed.
In a model-dependent search, all data during the period of gamma emission
reported by any satellite was examined, with the energy spectrum
predicted from gamma-ray spectra of individual GRBs.
The models tested were different formulations of the same fireball
phenomenology, producing neutrinos at proton-photon interactions in
internal shocks with their standard parameters and uncertainties on
those parameters.
The model-independent analysis searched more generically for neutrinos
on wider time scales, up to the limit of sensitivity to small numbers
of events at $\pm1$ day, or with different spectra.
As a result, an upper limit on the flux of UHE neutrinos
associated with GRBs was found to be at least a factor of 3.7 below
the predictions of the models considered.
It was thus concluded that GRBs are not the only source of UHECRs or
the efficiency of neutrino production is much lower than has
been predicted.


However, Li argued after~\cite{icecube-nus-2011a,icecube-nus-2011b} were
published that the theoretical prediction of neutrinos from GRBs by
IceCube overestimates the GRB neutrino flux by a factor $\sim5$
because they ignore both the energy dependence of the fraction of
proton energy transferred to charged pions and the radiative energy loss
of secondary pions and muons when calculating the normalization of the
neutrino flux~\cite{Li-nu_flux}.
This point of view was later supported by H\"ummer et~al.~\cite{hummer_etal-2012}.
They kept intact the astrophysical parameters of the fireball model used
for the analysis of the IceCube data made in~\cite{icecube-nus-2011a} but
included the full spectral dependencies of the proton and photon spectra,
the cooling of the secondaries, flavour mixing and additional multi-pion,
kaon, and neutron production channels for the neutrinos.
As a result, a significant deviation in the normalization of the
predicted neutrino flux of about an order of magnitude, with a very
different spectral shape peaking at slightly higher energies was found.
Different arguments were put forward by Dar to demonstrate that
the IceCube collaboration over-interpreted their results and they do
not exclude GRBs as the main source of UHECRs~\cite{dar-2012-nus}.


There is also a class of so called ``top-down'' models, which predict
UHECRs from the decay of super-heavy relic particles, see,
e.g.,~\cite{bhat-sigl-2000} for a review.
These models are currently disfavoured by the limits of the PAO on the photon
fraction~\cite{PAO-photons-2009}, and we do not discuss them here.

As we have seen above, the problem of the origin of UHECRs might be
far from its solution, and the existing experiments will not necessarily
provide data sufficient for obtaining the final answer.
That is why special attention is now paid to the forthcoming
orbital experiments TUS
aboard the Mikhail Lomonosov satellite~\cite{pan-2011} and JEM-EUSO aboard
the International Space Station~\cite{ebisu-2011}.
Recall that TUS is the first orbital detector of UHECRs. It
consists of the two main parts, a Fresnel-type segmented
mirror-concentrator of $\sim2$~m$^2$ area and a photo receiver built
of 256 photomultiplier tubes.
The field of view (FoV) of the detector equals $4.5^\circ$ and covers an
area of $(H\times0.16)^2$~km$^2$, where $H$ changes from 550~km to 350~km
in three years of operation.
The energy threshold of the detector is $\sim70$~EeV.

The TUS detector is to a large extent a pathfinder aimed to test the 
concept of registering UHE particles from space suggested by
Benson and Linsley more than 30 years ago~\cite{benson-linsley-1981}.
The JEM-EUSO mission is a much more advanced experiment with a
2.5~m telescope with a super-wide ($60^\circ$) FoV as the main instrument.
It is expected that the cumulative exposure of JEM-EUSO will be
of the order of $10^6$ km$^2$~sr~yr at 300~EeV, which is approximately an order
of magnitude greater than possibly achieved by the PAO.
With JEM-EUSO, baryons, photons and neutrino primaries can be discriminated
with considerable accuracy, and upper limits to the fluxes
of the last two will be improved by at least a factor of 10 beyond
present experiments.
The mass target inside the FoV is $\sim10^{12}$~ton, which depending
on the actual astrophysics scenario makes likely the observation
of up to a few cosmogenic neutrinos per year, thus providing an
additional opportunity to test the above models of sources of UHECRs.

\ack

The research was partially supported by Federal contract No.~16.518.11.7051.

\section*{References}
\bibliographystyle{iopart-num}
\bibliography{ecrs_pcr2_598}

\end{document}